\def\spose#1{\hbox to 0pt{#1\hss}}

\def\multleft#1{\hbox to size{\vbox {\halign {\lft{##}\cr #1}}\hfill}\par}
\def\multright#1{\hbox to size{\vbox {\halign {\rt{##}\cr #1}}\hfill}\par}
\def\Mdot{\hbox{$\dot M$}}
\def\degmark{^\circ}
\def\today{\ifcase\month\or January\or February\or March\or April\or May\or
      June\or July\or August\or September\or October\or November\or December\fi
      \space\number\day, \number\year}

\def\Lsun{\hbox{$\rm\thinspace L_{\odot}$}}

\def\Msun{\hbox{$\rm\thinspace M_{\odot}$}}

\newcommand{\Msolar}{\mbox{\,$\rm M_{\odot}$}}

\def\H2{\hbox{H$_{2}$}}

\newcommand{\gtsim}{\mbox{{\raisebox{-0.4ex}{$\stackrel{>}{{\scriptstyle\sim}}
$}}}}
\newcommand{\ltsim}{\mbox{{\raisebox{-0.4ex}{$\stackrel{<}{{\scriptstyle\sim}}
$}}}}
\documentstyle[epsfig]{mn}

\begin{document}
\hsize=6truein
       
\title[The cosmological evolution of quasar black-hole masses]
{The cosmological evolution of quasar black-hole masses}

\author[R.J.~McLure \& J.S.~Dunlop]
{Ross J. McLure$^{1}$\thanks{Email: rjm@roe.ac.uk}, James S. Dunlop$^{1}$\\
\footnotesize\\
$^{1}$Institute for Astronomy, University of Edinburgh, 
Royal Observatory, Edinburgh, EH9 3HJ, UK}
\maketitle
      
\begin{abstract}
Virial black-hole mass estimates are presented for 12698 quasars
in the redshift interval $0.1\leq z \leq 2.1$, based on modelling of 
spectra from the Sloan Digital Sky Survey (SDSS) first data release . 
The black-hole masses of the SDSS quasars are found to lie 
between $\simeq10^{7}\Msun$ and an upper
limit of $\simeq 3\times 10^{9}\Msun$, entirely consistent with the 
largest black-hole masses found to date in the local
Universe. The estimated Eddington ratios of the broad-line quasars (FWHM\,$\geq
2000$ km s$^{-1}$) show a clear upper boundary at 
$L_{bol}/L_{Edd}\simeq 1$, suggesting that the Eddington luminosity is
still a relevant physical limit to the accretion rate of 
luminous broad-line quasars at $z\leq 2$. By combining the 
black-hole mass distribution of the SDSS quasars with the 
2dF quasar luminosity function, the number density of 
active black holes at $z\simeq 2$ is estimated as a function of mass. 
In addition, we independently estimate the local black-hole mass 
function for early-type galaxies using the $M_{bh}-\sigma$ and 
$M_{bh}-L_{bulge}$ correlations. Based on the SDSS velocity dispersion 
function and the 2MASS $K-$band luminosity function, both estimates are
found to be consistent at the high-mass end ($M_{bh}\geq 10^{8}\Msun$). 
By comparing the estimated number density of active black holes 
at $z\simeq 2$ with the local mass density of dormant black holes, 
we set lower limits on the quasar lifetimes and find that the 
majority of black holes with mass $\geq 10^{8.5}\Msun$ are in place 
by $\simeq 2$. 
\end{abstract}


\section{INTRODUCTION}

The discovery that supermassive black holes are
ubiquitous among massive galaxies in the local Universe 
indicates that the majority of 
galaxies have passed through an active phase during their
evolutionary history. Moreover, the strong correlation observed between
black-hole and bulge mass (Kormendy \&
Richstone 1995; Magorrian et al. 1998; Gebhardt et al. 2000a; 
Ferrarese \& Merritt 2000) indicates that the evolution 
of the central black hole and its host galaxy are intimately related. 
Consequently, it is clear that studying the 
evolution of quasar black-hole masses will provide crucial 
information concerning the evolution of both quasars and massive 
early-type galaxies.

Within this context the last few years have seen renewed interest in the
possibilities of estimating the central black-hole masses of active
galactic nuclei (AGN). The major impetus for this has been the
results of the recent reverberation mapping programmes carried 
out on low-redshift quasars and Seyfert galaxies (Wandel, Peterson \&
Malkan 1999; Kaspi et al. 2000). The measurements of the 
broad-line region (BLR) radius produced by these long-term
monitoring programmes have allowed so-called virial black-hole mass
estimates to be made for 34 low-redshift AGN (Kaspi et al. 2000).  The 
principal assumption underlying the virial mass estimate 
is simply that the dynamics of the
BLR are dominated by the gravity of the central supermassive black
hole. Under this assumption an estimate of the central black-hole 
mass can be gained from : $M_{bh}\simeq G^{-1}R_{BLR}V_{g}^{2}$; where
$R_{BLR}$ is the radius of the BLR and $V_{g}$ is the velocity of the
line-emitting gas, as traditionally estimated from the 
FWHM of the $H\beta$ emission line. 

Given the large number of physical processes which could potentially
influence the dynamics of the BLR it is not immediately obvious that
the gravitational potential of the central black-hole should be dominant
(eg. Krolik 2001). However, several lines of evidence have recently
shown that the dynamics of the BLR appear to be at least consistent
with the virial assumption. Firstly, for the small
number of objects for which it is possible to do so, Peterson \&
Wandel (2000) have shown that the motions of the broad-line gas are
consistent with being virialized, with the velocity-widths of 
emission lines produced at different radii following the expected
$V\propto r^{-0.5}$ relationship. Secondly, the black-hole mass
estimates produced by the virial method are in good agreement
with the predictions of the tight correlation between black-hole mass
and stellar-velocity dispersion (Gebhardt et al. 2000b; Ferrarese et
al. 2001; Nelson et al. 2003; Green et al. 2003). Finally, using 
the virial estimator McLure \& Dunlop (2002) recently 
demonstrated, for 72 AGN at $z<0.5$, that AGN host galaxies 
follow the same correlation between black-hole mass and bulge 
mass as local quiescent galaxies. Viewed in isolation, each of 
these lines of evidence might be seen as simply a consistency check. 
However, taken together they offer good supporting evidence for 
the validity of the virial assumption.

Due to the fact that reverberation mapping measurements
of $R_{BLR}$ are necessarily reliant on high-accuracy monitoring
programmes lasting many years, at present such measurements are
only available for a small sample of low-redshift
AGN. Consequently, for estimating the black-hole masses for
large samples of AGN, a crucial result arising from the Kaspi et
al. (2000) study was that $R_{BLR}$ is strongly correlated with the
AGN monochromatic continuum luminosity at 5100\,\AA\,. By exploiting
this correlation it is therefore possible to produce a virial
black-hole mass estimate based purely on a luminosity and $H\beta$ FWHM
measurement. The availability of this technique has led to a
proliferation of studies of AGN black-hole masses in the recent 
literature.  These studies have primarily focused on 
low-redshift ($z\,\ltsim\, 0.5$) AGN samples, 
investigating the relationships between black-hole mass, the properties
of the surrounding host galaxies and the spectral energy 
distribution of the central engine (e.g. Laor 1998, 2000, 2001; Wandel 1999;
McLure \& Dunlop 2001,2002; Lacy et al 2001; Dunlop et al 2003;
Vestergaard 2004). However, the usefulness of the virial estimator 
based on the H$\beta$ emission
line is limited by the fact that $H\beta$ is redshifted out of
the optical at $z\geq0.8$. Consequently, the use of this emission line 
to trace BLR velocities in high-redshift AGN requires 
infra-red spectroscopy, which is relatively 
observationally expensive and limited to the available 
atmospheric transmission windows.

However, McLure \& Jarvis (2002) and Vestergaard (2002) have
recently demonstrated that this problem can be overcome by using the MgII and
C{\sc iv} emission lines as rest-frame UV proxies for $H\beta$. In addition, 
McLure \& Jarvis (2002) showed that $R_{BLR}$ is also strongly
correlated with the monochromatic continuum emission at 
3000\,\AA\,, allowing black-hole mass estimates for
high-redshift quasars to be made from a single optical spectrum
covering the MgII emission-line.

The availability of the new rest-frame UV black-hole mass estimators
has been exploited by recent studies to
investigate the black-hole masses of the most luminous quasars (Netzer
2003), the evolution of the black-hole mass - luminosity 
relation (Corbett et al. 2003) and also to estimate the mass of
the most distant known quasar at $z=6.41$ (Willott, McLure
\& Jarvis 2003; Barth et al. 2003). In particular, using composite spectra
generated from $\geq 22000$ 2dF+6dF quasars, Corbett et al (2003) 
successfully demonstrated that the evolution of the 
black-hole mass - luminosity relation is too weak to explain 
the evolution of the $z\leq 2.5$ quasar luminosity function with 
a single population of long-lived objects.

\begin{figure*}
\centerline{\epsfig{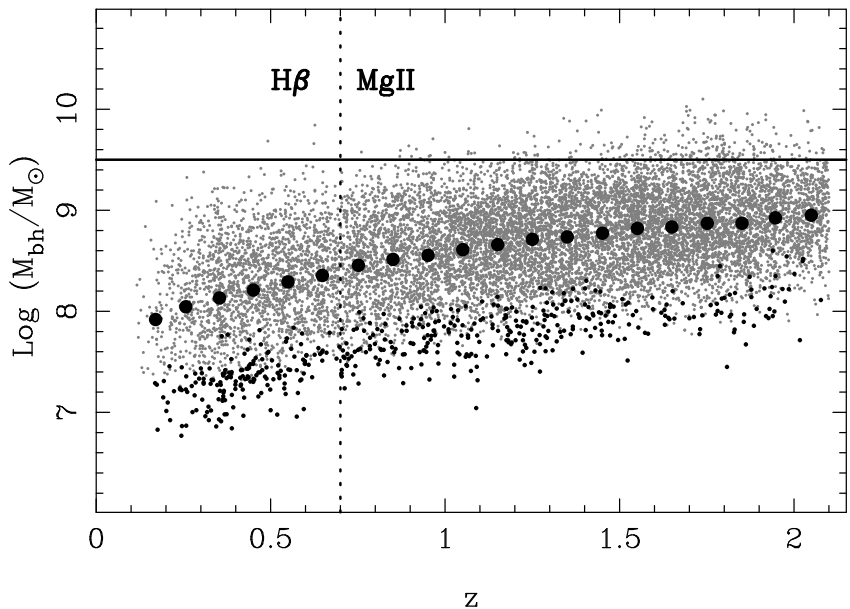}}
\caption{Virial black-hole mass estimate
versus redshift for our full SDSS quasar sample. Broad-line quasars
(FWHM$\,\geq 2000$ km s$^{-1}$) are shown as grey symbols, while
narrow-line objects (FWHM$<2000$ km s$^{-1}$) are shown as black
symbols. The mean black-hole 
masses within $\Delta z=0.1$ bins are shown as filled circles
(standard errors are smaller than the symbols except for 
lowest redshift bin). The vertical dotted line highlights 
the switch from using 
the $H\beta$-based to the MgII-based virial mass estimator at
$z=0.7$. The horizontal solid line marks a black-hole mass of
$3\times 10^{9}\Msun$, the maximum mass observed at low
redshift (see text for discussion).}
\label{fig1}
\end{figure*}
The recent publication of the SDSS first data release 
has provided publically available, fully calibrated, optical spectra 
of $\ge 17000$ quasars in the redshift interval $0.08<z<2.3$. The flux
calibration, long wavelength coverage (3800\,\AA $<\lambda <\,$9200\,\AA)
and good spectral resolution ($1.4$\,\AA/pix) of the SDSS spectra make
them ideal for use in virial black-hole mass estimates. In this paper 
we study the black-hole mass distribution of quasars in 
the redshift range $0.1<z<2.1$ by utilizing 
continuum and emission-line fits to SDSS quasar spectra, together
with calibrations of the virial mass estimators for $H\beta$ and MgII
based on those derived by McLure~\&~Jarvis (2002). Via
cross-calibration with low-redshift AGN with actual reverberation
mapping black-hole mass estimates (Kaspi et al. 2000), it is known 
that the scatter in the virial mass estimators based on both the
$H\beta$ and MgII emission lines is a factor of 2.5 
(see McLure \& Jarvis 2002; appendix) and, given the potential
uncertainties, it is probably unwise to trust individual virial
black-hole mass estimates to within a factor of $\simeq 3$. However, 
because there is no systematic off-set between the 
virial estimators and the reverberation 
mapping results, the virial mass estimators are expected to produce 
statistically accurate results when applied to large samples such as the SDSS.

The structure of the paper is as follows. In Section 2 we 
described the criteria used for selecting our sample
from the SDSS Quasar Catalogue II of Schneider et al. (2003). In Section
3 the basic distribution of the SDSS quasar 
black-hole masses as a function of redshift is presented. This new
information is then used to address important questions regarding
the virial black-hole mass estimator and the the accretion rate of
luminous quasars. Firstly, a recent study by Netzer (2003) based on the 
C{\sc iv} and H$\beta$ virial mass estimators has suggested that many 
luminous quasars at $z\,\gtsim\,2$ may harbour
central black holes with masses in excess of $10^{10}\Msun$. As
highlighted by Netzer (2003), since galaxies with suitably 
large velocity dispersions ($\sigma\,\gtsim\, 600$ km s$^{-1}$) are not 
observed at low redshift, this result raises questions about either
the reliability of the virial mass estimator, or the form of the
bulge:black-hole mass relation, at high redshift. We re-address this 
question using our new SDSS quasar black-hole mass estimates 
based on the MgII calibration of the virial mass estimator. 

Secondly, recent studies have shown that at 
low-redshift ($z\leq0.5$), luminous broad-line quasars are 
accreting below the Eddington limit (Floyd et al. 2003; Dunlop et
al. 2003; McLeod \& McLeod 2001). However, it has been 
suggested (eg. Mathur 2000) that narrow-line
Seyfert galaxies (NLS1) are the low-redshift analogs of the early
evolution of luminous quasars, where the central black hole is still
in an exponential growth phase, accreting beyond the Eddington limit.
Consequently, if NLS1 are the low-redshift analogs of the early 
evolution of luminous quasars, then
at high redshift we might expect to see luminous quasars accreting
above the Eddington limit. Using the black-hole mass and
bolometric luminosity estimates for our SDSS sample, we investigate
whether the Eddington luminosity is still a physically relevant limit to
quasar accretion rates at $z\leq 2$. 

In Section 4 we re-investigate 
the form of the local black-hole mass
function within the context of the latest results on 
the $M_{bh}~-~L_{bulge}$ and $M_{bh}~-~\sigma$ correlations. In Section 5
this information is combined with the new SDSS quasar 
black-hole masses and the 2dF quasar
luminosity function to estimate the activation fraction of $z\simeq 2$
black holes as a function of mass, and deduce lower limits on the
lifetimes of the most luminous quasars. In Section 6 we summarize 
our main results and conclusions. Full details of the calibration of the
virial black-hole mass estimators and the emission-line modelling of
the SDSS quasar spectra are provided in the Appendix. Unless 
otherwise stated all calculations assume the following 
cosmology : $H_{0}~=70$~km~s$^{-1}$Mpc$^{-1}$, $\Omega_{m}=0.3$,
$\Omega_{\Lambda}=0.7$. 

\section{The sample}

The sample of quasars analysed in this paper is drawn from the
SDSS Quasar Catalog II (Schneider et al. 2003). This catalog is 
based on the SDSS first data release, and
features 16713 quasars with $M_{i}(AB)<-22$ in the redshift interval 
$0.08~<~z~<~5.41$. The first stage in constructing our sample was to
select all objects from the Schneider et al. catalog in the redshift
range $0.08~<~z~<~2.1$, with the upper redshift limit imposed to ensure 
sufficient wavelength coverage to reliably determine the 
MgII emission-line widths. The resulting list of 14181 quasars was then
passed through the emission-line modelling software described 
in the Appendix.

The final sample used in the analysis consists of 12698 quasars, or 
90\% of the $z<2.1$ quasars in the Schneider et al. catalogue. The 
remaining 10\% of objects 
were excluded for having unreliable modelling results due to 
low signal-to-noise, 
particularly weak emission lines or having been affected by 
artifacts in the spectra. In 
conclusion, although the SDSS Quasar Catalog II is not 
statistically complete (Schneider et al. 2003), our final sample of 
12698 quasars is clearly representative of optically luminous quasars 
in the redshift interval $0.1<z<2.1$.

\section{The evolution of quasar black-hole masses}

In Fig \ref{fig1} we show the distribution
of virial black-hole mass estimates versus redshift for our 
full SDSS quasar sample. Broad-line quasars
(either H$\beta$ or MgII emission line with FWHM$\,\geq2000$ km
s$^{-1}$) are shown as grey symbols and narrow-line objects (FWHM$<2000$ km
s$^{-1}$) are shown as black symbols. The mean black-hole masses within
$\Delta z=0.1$ bins are shown as filled circles, and the vertical
dotted line at $z=0.7$ marks the transition between the use of the
$H\beta$ and MgII based virial estimators. There can be seen to be a
smooth transition between the two virial mass calibrations, and more
details concerning the quality of the agreement between the $H\beta$
and MgII based virial estimators are provided in the Appendix. Finally,
the solid horizontal line is at a black-hole mass of $3\times
10^{9}\Msun$, which is representative of the maximum mass which has
been dynamically measured within the local galaxy population (Ford et 
al. 1994; Tadhunter et al. 2003). 

\subsection{Black-hole mass as a function of redshift}

\begin{figure*}
\centerline{\epsfig{file=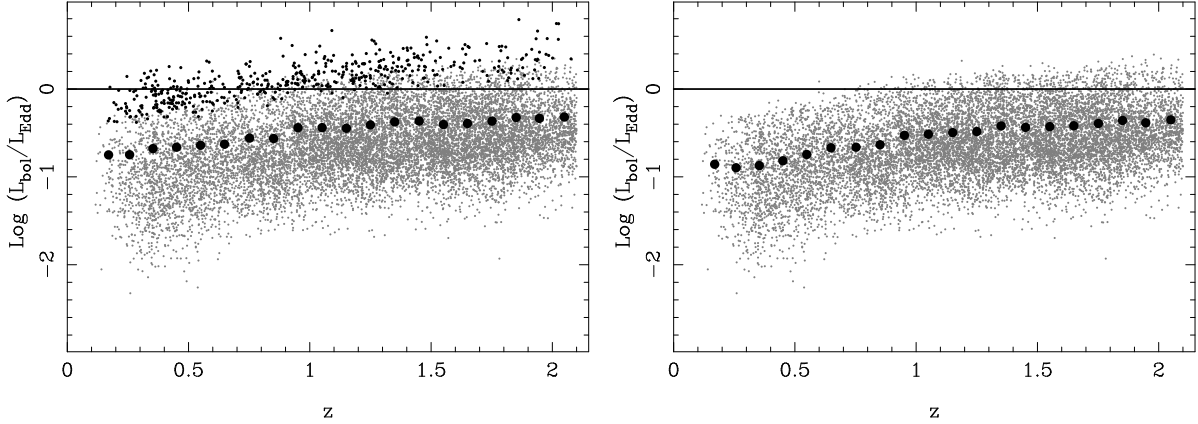,width=16.0cm,angle=0}}
\caption{Panel A shows
bolometric luminosity as a fraction of the Eddington limit versus
redshift for the
full sample. The grey symbols are broad-line quasars 
(FWHM$\,\geq2000$ km s$^{-1}$) and the black symbols are 
narrow-line objects (FWHM$<2000$ km s$^{-1}$). The mean values 
of $L_{bol}/L_{Edd}$ within $\Delta z=0.1$ bins are shown as filled circles. 
The solid horizontal line marks the Eddington limit. Panel B is
identical to panel A except that only the broad-line 
quasars have been plotted.}
\label{fig2}
\end{figure*}
The mean black-hole mass can be seen to increase as a
function of redshift, rising from $\simeq 10^{8}\Msun$ at $z\simeq
0.2$ to $\simeq 10^{9}\Msun$ at $z\simeq 2.0$. However, this is
entirely as expected due to a combination of the effective flux-limit
of the sample and the r\^{o}le of the quasar continuum luminosity 
within the virial mass estimator. In the event that the mean FWHM 
remains approximately constant with
redshift, the mean virial black-hole mass estimate is required to
increase as $<\lambda L_{\lambda}>^{0.6}$ by construction
(Eqn A6 \& Eqn A7). The increase in mean black-hole mass with
redshift seen in Fig \ref{fig1} is entirely consistent with
this effect. 

\subsection{Evidence for a limiting black-hole mass}

Amongst the sample of local galaxies for which the central black-hole
mass has been measured via gas or stellar dynamics, the most massive
black holes which have been found to date are in M87 (Ford et al. 1994) and 
Cyg A (Tadhunter et al. 2003), both with $M_{bh}\simeq 3\times
10^{9}\Msun$.

Given the relatively small cosmological volume from which this sample
of local galaxies is drawn, it is of interest to investigate whether
$\simeq~3\times~10^{9}~\Msun$ represents a genuine 
limiting mass. Furthermore, the
$M_{bh}-\sigma$ and $M_{bh}-L_{bulge}$ correlations, defined by the
local galaxies with dynamical black-hole mass measurements, make
definite predictions for what the maximum black-hole mass is likely to
be. For example, the distribution of stellar-velocity dispersions for
early-type galaxies from the SDSS (Bernardi et al. 2003; Seth et
al. 2003) show that velocity dispersions of $\gtsim$\,400 km s$^{-1}$ 
are exceptionally rare among the local galaxy population. Combined with the 
latest determination of the $M_{bh}-\sigma$
relation (Tremaine et al. 2002), this implies a limiting black-hole 
mass of $\simeq 2 \times
10^{9}\Msun$, in excellent agreement with the largest black-hole
masses measured to date. Consequently, if the virial black-hole mass
estimator predicts that a large population of luminous quasars at high
redshift harbour black holes with masses $\geq 10^{10}\Msun$, the
implied velocity dispersions of $\sigma \simeq 600$
km s$^{-1}$ would be in direct conflict with the known properties of 
early-type galaxies.

This issue was recently addressed by Netzer (2003), who
compiled virial black-hole mass estimates (based on $H\beta$ and C{\sc iv}
emission-line widths) for a heterogeneous sample of 724 quasars in the
redshift interval $0<z<3.5$. The results of this study indicated that a
significant number of luminous ($\lambda L_{1350}\geq
10^{46.5}$ergs\,s$^{-1}$) quasars are  powered by black holes with
masses~$\gtsim 10^{10}\Msun$. Consequently, Netzer (2003) concluded that
either the virial mass estimator (principally the 
$R_{BLR}-\lambda L_{\lambda}$ correlation) was invalid at
$z\,\gtsim\,2$ or, alternatively, that the relationship between black-hole and
bulge mass has a different form at high redshift, presumably because 
the quasar hosts are still in the process of forming.
 
However, the new results presented here for the SDSS quasars do not
appear to support this conclusion. It can be seen from Fig
\ref{fig1} that the upper envelope of black-hole masses for the
SDSS quasars is fully consistent with a maximum of 
$\simeq 3\times 10^{9}\Msun$, with a negligible number of 
quasars exceeding this limit at $z\leq 1.5$. At
the high-redshift end of the sample ($1.5<z<2.1$) there are only a small
number of quasars with virial black-hole mass estimates $\geq 3\times
10^{9}\Msun$, less than 5\% of the SDSS 
quasars with $z\geq 1.5$, and only two objects are found to have virial
black-hole mass estimates in excess of $10^{10}\Msun$. When the level
of scatter associated with the virial mass estimator ($\simeq 0.4$ dex)
is taken into account, it is clear that there is no conflict between
the SDSS quasar black-hole masses and either the locally observed
maximum black-hole mass or, perhaps more significantly, the 
maximum black-hole mass predicted from combining the 
stellar-velocity dispersion distribution of early-type 
galaxies (Bernardi et al. 2003) with the $M_{bh}-\sigma$ relation. 

The reason for the discrepancy between the new results presented here
and those of Netzer (2003), hereafter N03, is not entirely clear. 
The distribution of black-hole mass as a function of redshift 
determined by N03 shows an upper limit consistent with $M_{bh}\simeq
10^{10}\Msun$, with a small number of objects with estimated masses
exceeding this limit. In contrast, the results for the SDSS quasars
presented in Fig \ref{fig1} are consistent with a limit of
$\simeq 3\times 10^{9}\Msun$, a difference of 0.5 dex. The discrepancy 
cannot be due to the use of different calibrations 
of the $R_{BLR}-\lambda L_{\lambda}$ correlation. If fact, if our
virial mass estimator is adjusted to the same 
$R_{BLR}-\lambda L_{\lambda}$ slope ($\lambda L_{\lambda}^{0.58}$) 
and the same geometric normalization (N03 adopt $f=\sqrt{3}/2$, 
we adopt $f=1$, see Eqn A1) our black-hole mass estimates for the most
luminous SDSS quasars are actually reduced by $\simeq 0.2$
dex. Presumably the cause of the discrepancy must therefore lie either
in the different methods used to determine the 
quasar continuum luminosities, or N03's use of C{\sc iv} as a 
proxy for $H\beta$ instead of the MgII FWHM measurements adopted here.
In this context we note that Vestergaard (2004) recently found
evidence for a similar limiting black-hole mass to that found here,
using a smaller sample of $2\,\ltsim\, z\, \ltsim\, 4$ quasars and 
the C{\sc iv}-based virial mass estimator.

In conclusion, based on the $H\beta$ and MgII virial black-hole 
mass estimates for $>12000$ SDSS quasars presented here, we find no 
evidence that $z\leq 2$ quasars harbour central black-holes with masses 
in excess of the maximum of  $\simeq 3 \times 10^{9}\Msun$ 
observed in the local Universe. This consistency between 
the most massive black-holes at $z\simeq 2$ and the 
most massive black-holes at $z\simeq 0$ is
important given that it is known from low-redshift studies that the
host galaxies of these $z\simeq 2$ quasars must be, or at least evolve
into, massive early-type galaxies (Dunlop et al. 2003; McLure \&
Dunlop 2002; McLeod \& McLeod 2001). Reassuringly, the black-hole mass
distribution of the SDSS quasars presented here is entirely consistent
with their host galaxies having similar properties to those of
low-redshift early-types; i.e. $\sigma\,\ltsim\, 400$~km s$^{-1}$ and 
$L\,\ltsim \,10\, L^{\star}$.

\subsection{Quasar accretion rates}
\begin{figure}
\centerline{\epsfig{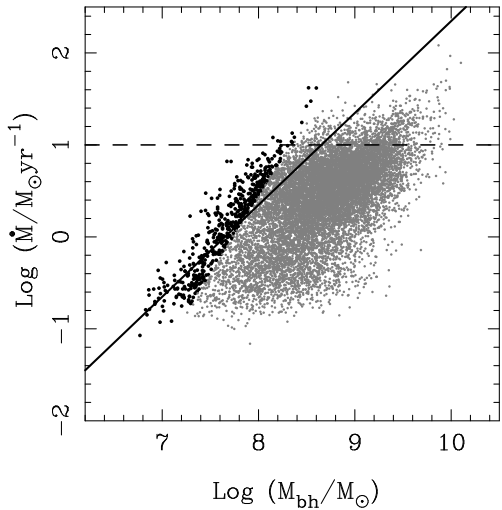}}
\caption{The distribution of the full SDSS sample on the accretion 
rate $-$ black-hole mass plane. The grey
symbols are broad-line quasars (FWHM$\,\geq2000$ km s$^{-1}$) and the
black symbols are narrow-line objects (FWHM$<2000$ km s$^{-1}$). The 
solid line shows the location of the Eddington limit
($L_{bol}=L_{Edd}$), while the dashed line highlights an accretion
rate of ten solar masses per year, below which 94\% of the sample lie.}
\label{fig3}
\end{figure}

In panel A of Fig \ref{fig2} we 
plot the distribution of quasar Eddington ratios ($L_{bol}/L_{Edd}$)
as a function of redshift for the full sample (narrow + broad-line
objects). The mean $L_{bol}/L_{Edd}$ values within $\Delta z=0.1$ bins 
are shown as the filled circles, with the solid horizontal line 
highlighting the Eddington limit ($L_{bol}/L_{Edd}=1$). Panel B is
identical to panel A except that only the broad-line objects 
are plotted. Several features of the distribution of the SDSS quasars 
on the $L_{bol}/L_{Edd}-z$ plane are worthy of individual 
comment and are discussed in turn below.

In common with the distribution of black-hole masses, 
it can be seen that the mean Eddington ratio 
of the SDSS quasars increases with redshift, rising from
$L_{bol}\simeq 0.15 L_{Edd}$ at $z\simeq 0.2$ to $L_{bol}\simeq 0.5
L_{Edd}$ at $z\simeq 2.0$ (details of how bolometric luminosities have been 
estimated from the quasar continuum luminosities are provided 
in the Appendix). However, due to the fact that the Eddington luminosity
is directly proportional to the estimated black-hole mass, the
mean Eddington ratio is required to scale as
$L_{bol}/L_{Edd}\propto L_{bol}^{\sim 0.4}$ in the event that 
the mean FWHM is approximately constant with redshift. The increase in
the mean Eddington ratio with redshift seen in Fig \ref{fig2} is 
entirely consistent with this expectation.

The distribution of Eddington ratios displayed by the SDSS quasars
clearly shows that assuming that all luminous quasars are accreting
at their Eddington limit is a poor approximation. This 
result is important because it is often assumed that optically luminous
quasars are accreting at their Eddington limit within models of quasar
evolution. In fact, our results show that a constant accretion 
rate of $\simeq 0.3\rightarrow 0.4  L_{Edd}$ is a much 
better approximation than using the Eddington luminosity. 

In panel A of Fig \ref{fig2} it can be seen that the majority of the
narrow-line objects (black symbols) are estimated to be accreting at
super-Eddington rates. This result is in agreement with numerous
studies of low-redshift narrow-line Seyfert 1 galaxies (NLS1) which, in the
most widely accepted model, are thought to be examples of 
low black-hole mass/high accretion rate AGN. However, it is also 
important to remember that the narrow-line portion of the 
SDSS quasar sample is likely to be 
an admixture of genuine narrow-line objects 
(i.e. low black-hole mass/high accretion rate) and intrinsically 
broad-line quasars which appear as narrow-line objects simply because 
they are viewed close to pole-on. Indeed, as discussed in the appendix,
if there is any significant disk-component to the FWHM of the broad
emission lines then some level of contamination by pole-on sources 
is inevitable. Within
this context it is noteworthy that we do not see any trend for 
an increase in the numbers of narrow-line objects with 
redshift, as might be expected under the hypothesis that NLS1s are 
the low-redshift analogs of the early evolutionary stages of 
luminous broad-line quasars (eg. Mathur 2000). On the contrary, the 
fraction of narrow-line objects in our SDSS sample decreases 
from $\simeq 10\%$ at $z\simeq 0.2$ to only $\simeq 1\%$ at 
$z\simeq 2$ (see Fig \ref{fig1}). In fact, this trend is at 
least qualitatively consistent with 
the predictions of a simplified model of the BLR which depends on both
orientation and luminosity.  Firstly, if it is assumed that 
all quasars are intrinsically 
broad-line objects, and that the BLR has a flattened geometry, then 
narrow-line objects are simply those which are 
viewed close to pole-on ($\leq 15\degmark$). In a flux-limited sample
the proportional of narrow-line objects is then predicted to fall with
redshift if the opening angle of the obscuring torus is dependent 
on the AGN luminosity; i.e $\theta \propto L^{0.5}$ (the 
so-called ``receding torus'' model e.g. Simpson 1998). 
In this simplified scheme the redshift evolution of the narrow-line fraction 
seen in Fig \ref{fig1} is roughly consistent with a receding torus 
model with a half-opening angle of $\simeq
45\degmark$ at the low-redshift/luminosity end of the SDSS sample, 
rising to $\simeq 80\degmark$ at the high-redshift/luminosity end.

It can be seen from panel B of Fig \ref{fig2} that at 
redshifts $1<z<2$ the upper envelope of the Eddington ratios for the
{\it broad}-line quasars is virtually flat at $L_{bol}/L_{Edd}\simeq 1$. 
Indeed, the fact that we do not see significant numbers of quasars 
which are accreting at greater than the Eddington limit, with virtually none
accreting at $\geq 2 L_{Edd}$, can be taken as strong evidence
that the Eddington limit is a physically meaningful upper limit to 
the accretion rate in broad-line quasars, at least at $z\,\ltsim\, 2\,$.

In Fig 3 we show the distribution of the SDSS
quasars on the accretion rate $-$ black-hole mass plane, where the
accretion rates have been calculated assuming a canonical 
mass-to-energy conversion efficiency of $\epsilon=0.1$. The solid line 
in Fig \ref{fig3} shows the location of the Eddington limit 
($L_{bol}=L_{Edd}$), while the dashed line highlights an accretion 
rate of ten solar masses per year. It can immediately be seen 
from Fig \ref{fig3} that the vast majority (94\%)
of the SDSS quasars are accreting material at rates of 
$\Msolar~<~\Mdot~<~10\Msolar$ per year. However, perhaps the 
most striking feature of Fig 3 is the near total lack of quasars with
black-hole masses of $M_{bh}>10^{9}\Msun$ and accretion rates close to
the Eddington limit. This feature must be regarded as significant
because it is extremely unlikely that large numbers of such objects
are missing from the SDSS quasar sample. We note here that this 
situation is at least qualitatively consistent with models of black-hole
growth in which the exponential, Eddington-limited, growth of the
black-hole is terminated by a physical limit to the amount of material
which can be supplied for accretion (eg. Archibald
et al. 2002; Granato et al. 2003, 2001). In models of this type,
following the end of the exponential-growth phase the black hole is
free to remain active as a luminous quasar for $\simeq 10^{8}$ years, accreting
below the Eddington limit, without leading to the production of large
numbers of black holes with masses $>10^{10}\Msun$ which are not
observed locally. This type of scenario is also consistent with 
the results shown in Fig 1 \& Fig 2, and is discussed in more 
detail in Section 5.

\section{The local mass function of dormant black holes}

\begin{figure}
\centerline{\epsfig{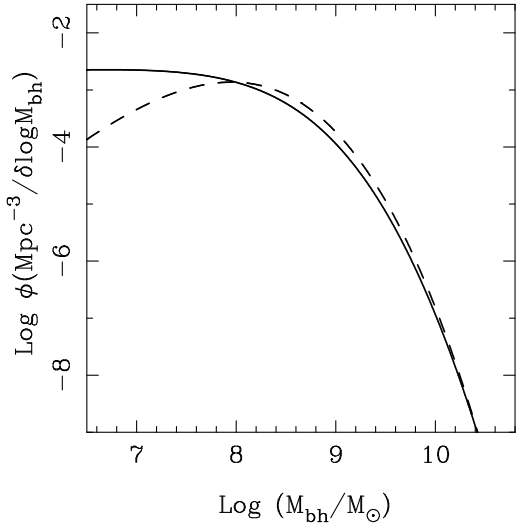}}
\caption{Two independent estimates of the redshift zero dormant black-hole mass
function for early-type galaxies. The solid line is a transformation 
of the $K-$band early-type  galaxy luminosity function of 
Kochanek et al. (2001) using the $M_{bh}-L_{bulge}$ correlation of 
McLure \& Dunlop (2002). The dashed line is a transformation of 
the SDSS early-type galaxy stellar-velocity dispersion 
function of Seth et al. (2003) using the Tremaine et al. (2002) 
fit to the $M_{bh}-\sigma$ correlation.}
\label{fig4}
\end{figure}

Following the discovery that supermassive black holes appear 
to be ubiquitous at the centres of massive local galaxies, 
it has been of interest to use this information to 
estimate the mass function of dormant black holes 
in the local Universe. From a theoretical perspective it was 
shown by Soltan (1982) that the local mass density of 
dormant black holes could be calculated from the
total radiated energy of optical quasars, which in turn can be
calculated from the quasar source-count distribution. Traditionally 
such estimates predict local black-hole mass densities 
a factor of a few lower (Soltan 1982; Chokshi \& Turner 1992; Yu \& Tremaine
2002) than those estimated from the X-ray background 
(eg. Fabian \& Iwasawa 1999), suggesting that the 
dominant fraction of the quasar population is optically obscured. 
Given that a principal objective of quasar evolution models 
is to simultaneously explain the evolution of both the 
quasar luminosity and black-hole mass functions, it is clearly
important to have a robust measurement of the form of the local 
mass function of dormant black holes.

Previous estimates of the local black-hole
mass function in the literature have relied on either the $M_{bh}-L_{bulge}$
correlation to make the transformation from the local galaxy
luminosity function (e.g. Salucci et al. 1999), or the correlation
between radio luminosity and black-hole mass observed in local 
early-type galaxies (e.g. Franceschini et al. 1998). More recently,
both Yu \& Tremaine (2002) and Aller \& Richstone (2002) exploited the
tight $M_{bh}-\sigma$ correlation to derive an estimate of the local
mass density of dormant black holes. While the Yu \& Tremaine (2002) 
calculation was based directly on the SDSS stellar-velocity dispersion 
function, the Aller \& Richstone (2002) calculation used the 
Faber-Jackson relation (Faber \& Jackson 1976) to first estimate the 
dispersion function from the galaxy luminosity function. Despite
this difference in approach, both studies found consistent 
results for the local black-hole mass density (early+late types) 
with estimates  of $(2.4\pm 0.8) \times 10^{5}\Msun$Mpc$^{-3}$ and 
$(2.9\pm 0.5) \times 10^{5}\Msun$Mpc$^{-3}$ for Aller \& Richstone and Yu \&
Tremaine respectively ($H_{0}=70$ km s$^{-1}$Mpc$^{-1}$).

In Section 5 we will proceed to estimate the activation fraction of
supermassive black holes at $z\simeq 2$. This calculation requires
a knowledge of the actual functional form of the local dormant 
black-hole mass function, information which is unavailable from the
previous studies in the literature. At $z\simeq 2$ we are 
exclusively interested in the form of high-mass end of the local
black-hole mass function, and consequently, in this section 
we use the $M_{bh}-\sigma$ and $M_{bh}-L_{bulge}$ relations to 
derive two independent estimates of the local black-hole mass 
function for early-type galaxies.

\subsection{The velocity dispersion estimate}

Seth et al. (2003) recently performed a detailed analysis of the
stellar-velocity dispersion function of 9000 early-type galaxies drawn
from the SDSS. As part of this analysis Seth et al. provide a fitting
formula for the dispersion function, similar to a Schechter
function in form, which is governed by four free parameters
$(\phi_{\star}, \sigma_{\star},\alpha,\beta)$. As the starting point
for our calculation we adopt the best fit determined by Seth et al. to the
observed dispersion function, including measurement errors, which is
described by the parameter values (0.002 Mpc$^{-3}$, 88 km s$^{-1}$,
6.5, 1.8). Secondly, we then apply a change of variables 
to convert the velocity dispersion function into a black-hole mass
function using the latest version of the $M_{bh}-\sigma$ correlation
($M_{bh} \propto \sigma^{4.02\pm0.32}$) as derived by Tremaine et
al. (2002). The final step in the process is to convolve this estimate
of the black-hole mass function with a gaussian of $\sigma=0.27$~dex
to account for the associated scatter of the $M_{bh}-\sigma$
relation. The estimate of the local black-hole mass function
produced via this method is shown as the dashed line in Fig
\ref{fig4}.

\subsection{The bulge luminosity estimate}

Our second approach involves a conversion of the local galaxy
luminosity function into an estimate of the local black-hole mass
function using the $M_{bh}-L_{bulge}$ correlation. This approach 
was previously used by Salucci et al. (1999), who
derived an estimate of the local black-hole mass function via a
two-step process. The first stage was the construction of the local
bulge/spheroid mass function from representative optical galaxy
luminosity functions, segregated by morphological type, using an
appropriate mass-to-light ratio and an estimate for each morphological
type of the bulge:total light ratio. The second stage was to then 
convert the spheroid mass function into the black-hole mass function by
convolution with a log gaussian $M_{bh}/M_{bulge}$ distribution,
centred on $<M_{bh}/M_{bulge}>=-2.6$ with $\sigma=0.3$ dex. The 
Salucci et al. (1999) study demonstrated that at the 
high-mass end ($M_{bh}\geq 10^{8}\Msun$), the local dormant 
black-hole mass function should be entirely dominated by the 
contribution from early-type (E+S0) galaxies.

We are now in a position to re-visit this calculation using improved
information on the luminosity function of early-type galaxies and the
on form of the black-hole mass - bulge mass relation. For the purposes of 
our estimate we adopt the 2MASS early-type $K-$band luminosity function of 
Kochanek et al. (2001). The Kochanek et al. early-type luminosity 
function is conveniently segregated to included only S0 and 
elliptical galaxies, making it well matched to the SDSS velocity 
dispersion function of Seth et al. (2003). Moreover,
due to the $K-$band selection, we can be confident that the galaxy
luminosities trace the mass of the old stellar spheroid population, 
which has been shown recently 
to correlate with central black-hole mass with a 
comparable scatter to the $M_{bh}-\sigma$ correlation; i.e $\simeq
0.3$ dex (McLure \& Dunlop 2002; Erwin et al. 2003; Marconi \& Hunt 2003). 

Specifically, we adopt the $K-$band luminosity function parameters 
derived by Kochanek et al. (2001) for their early-type (E+S0) sample:
$\alpha=-0.92$, $\phi=0.45\times 10^{-2} h_{100}^{3}$ Mpc$^{-3}$ and
$M_{K}^{\star}=-23.53 +5\log h_{100}$. After conversion of the
luminosity function parameters to our adopted cosmology, the value of
$M_{K}^{\star}$ was further correct by $\Delta m =-0.2$ to transform from
isophotal to total magnitudes (Kochanek et al. 2001). In order to make
the transformation from the luminosity function to the black-hole mass
function we adopt the best-fit to the $M_{bh}-L_{bulge}$ correlation
from McLure \& Dunlop (2002), which was determined from a sample of
exclusively elliptical galaxies with reliable dynamical black-hole
mass measurements. After accounting for the change in
cosmology, this relation becomes:
\begin{equation}
\log M_{bh} = 1.25(\pm0.05)\log\frac{L_{K}}{\Lsun} -5.76(\pm0.53)
\end{equation}
\noindent
where an average colour of $R-K=2.7$ for early-type E/S0 galaxies has
been adopted (GISSEL98 models; Bruzual \& Charlot 1993), and we 
have also assumed for this simplified calculation that the $K-$band 
bulge:total luminosity ratio for both elliptical and S0 galaxies 
is unity. Following
convolution with a gaussian with $\sigma=0.3$ dex to account for the
scatter around the  $M_{bh}-L_{bulge}$ correlation, the final estimate
of the early-type dormant black-hole mass function is shown as the
solid line in Fig \ref{fig4}.

\subsection{Integrated black-hole mass density}

It can be seen from Fig \ref{fig4} that the two independent
estimates of the local early-type dormant black-hole mass function are
in good agreement at the high-mass end ($M_{bh}\geq
10^{8}\Msun$). The turn-over of the velocity-dispersion based estimate
at low masses is an artifact of the SDSS velocity dispersion 
function, which turns over at $\sigma\,\ltsim\, 125$ km s$^{-1}$ due 
to the faint magnitude limit of the SDSS early-type 
sample (Bernardi et al. 2003). Finally, if we integrate our 
estimate of the local black-hole
mass function based on the $K-$band luminosity function, our estimate
of the local black-hole mass density in early type galaxies is:
\begin{equation}
\rho_{bh}=(2.8\pm 0.4)\times 10^{5} \Msolar {\rm Mpc}^{-3}
\end{equation}
\noindent
which, considering the potential uncertainties and simplified nature
of this calculation, can be seen to be in good agreement with the 
values determined by both Yu \& Tremaine (2002) and Aller \& Richstone (2002). 

The good agreement between the two estimates of the 
local dormant black-hole mass function found here is apparently 
in contrast to one of the findings of Yu \& Tremaine (2002), who found that 
using the $M_{bh}-L_{bulge}$ relation led to a black-hole mass
density estimate more than a factor of two greater than their velocity
dispersion estimate. Although small differences are introduced by our
different choice of $M_{bh}-L_{bulge}$ relation and galaxy luminosity
function, we note here that the principal source of this apparent 
discrepancy is that we have adopted virtually identical 
levels of scatter in the $M_{bh}-\sigma$ and $M_{bh}-L_{bulge}$
relations ($\simeq 0.3$ dex). In contrast, Yu \& Tremaine adopted 
a scatter of $0.5$ dex in the $M_{bh}-L_{bulge}$ relation, based on 
the fit to $B-$band data by Kormendy \& Gebhardt (2001). If the Yu \&
Tremaine $M_{bh}-L_{bulge}$ density estimate is adjusted to 
reflect a scatter of $\simeq 0.3$ dex then their two
density estimates become consistent, differing by less than
50\%. Although the relative levels of scatter associated with the
$M_{bh}-\sigma$ and $M_{bh}-L_{bulge}$ relations is currently a
question of some debate, our choice of adopting a low scatter of $\simeq
0.3$ dex in the $M_{bh}-L_{bulge}$ relation is justified, 
especially for early-types, by the recent results of several 
detailed studies (McLure \& Dunlop 2002; Erwin et al. 2003; 
Marconi \& Hunt 2003).

\section{Constraining the activation fraction of supermassive black holes}
\begin{figure*}
\centerline{\epsfig{file=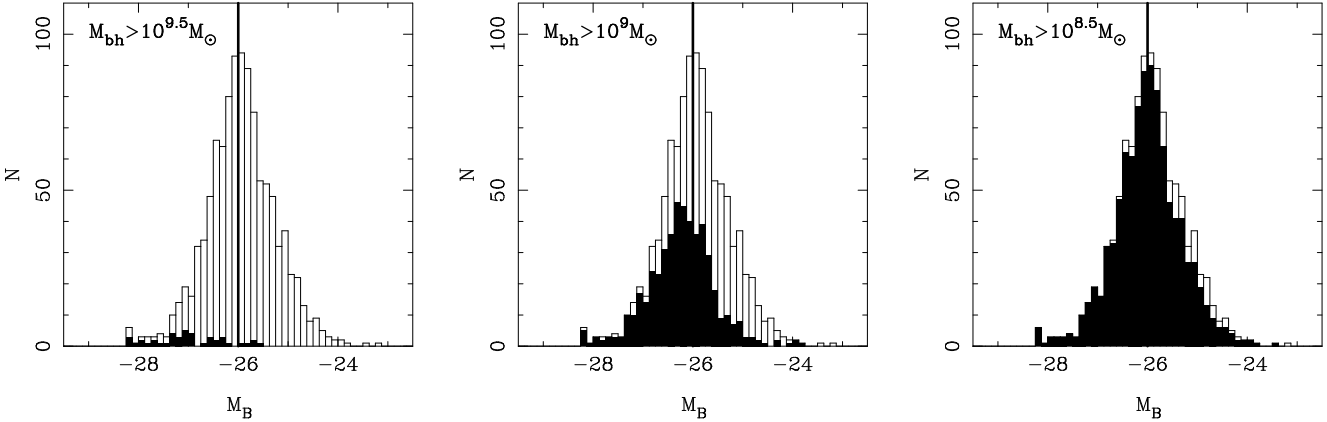,width=18.0cm,angle=0}}
\caption{Each panel shows the distribution of absolute $B-$band 
magnitude for the 1069 quasars from our SDSS sample in the 
redshift interval $1.90~<~z~<~2.1$. The black histograms show the 
distribution of sub-sets of these quasars whose black-holes are 
more massive than the thresholds shown in the top-left corner of each panel.}
\label{fig5}
\end{figure*}

Although the completeness of the SDSS Quasar Catalog II has not been
fully quantified (Schneider et al. 2003), there is no obvious reason 
to suspect that the SDSS quasars are not at least 
representative of luminous optical quasars. Therefore, it is possible
to use our new SDSS quasar black-hole mass estimates to study the
distribution of quasar black-hole masses as a function of luminosity.
Furthermore, by using the optical quasar luminosity function to 
calculate the number density of quasars brighter than a 
given absolute magnitude, it is then possible to estimate the 
number density of {\it active} black-holes as a function of mass. 
In this section we perform this calculation using our black-hole mass 
estimates for the SDSS quasars at $z\simeq 2$.

For the purposes of this calculation we adopt the quasar luminosity
function as determined by Boyle et al. (2000) for their combined
2dF+LBQS sample of more than 6000 quasars in the redshift range
$0.35<z<2.30$. The Boyle et al. luminosity function is defined in
terms of absolute $B-$band magnitudes, so we initially convert the 
bolometric luminosity estimates for our SDSS quasars into absolute
$B-$band magnitudes, using a transformation defined by $>300$ quasars
common to both the SDSS and the 2dF 10K quasar catalog 
(Croom et al. 2001; see Appendix). 

In Fig \ref{fig5} we show the distribution of 
absolute $B-$band magnitude for 1069 SDSS quasars in the redshift 
interval $1.90~\leq~z~\leq~2.10$. The filled histograms in each panel 
show the distribution of absolute $B-$band
magnitudes for three sub-sets of these objects which have black-hole masses
greater than our chosen mass thresholds of $10^{8.5}\Msun, 
10^{9}\Msun$ \,\&\, $10^{9.5}\Msun$. The solid line at $M_{B}=-26$
shows the location of the $m_{i}(AB)=19.1$ (Schneider et al. 2003)
flux limit of the main SDSS multi-colour quasar selection 
algorithm at $z=2.1$. Provided that at a given quasar luminosity 
the SDSS is not significantly biased towards including only the 
most massive black-holes, the ratio of the filled to un-filled
histograms in Fig \ref{fig5} provides a method for determining lower
limits on the number densities of active black holes at $z\simeq 2$.

We perform this calculation in two steps. Firstly, for each of the
three mass thresholds we calculate the fraction $f$ of quasars with
black-hole masses above the threshold, in three magnitude ranges:
$-25<M_{B}<-26$, $-26<M_{B}<-27$ \& $-27<M_{B}<-28.3$ (switching to
0.5 magnitude bins makes little difference to the final estimate). The 
lower limit of $M_{B}=-25$ has been chosen to 
conservatively ensure that we still
have sufficient numbers of quasars in the filled histograms to
determine $f$ in each magnitude bin to better than 10\% accuracy. 
Although $M_{B}=-25$ is one
magnitude fainter than the SDSS quasar selection algorithm flux limit
at $z\simeq 2$, we are making the explicit assumption that the ratio
of the filled to un-filled histograms would remain unchanged, even in
the event that the SDSS quasar selection was complete to $M_{B}=-25$.
The second step in the process is to use the Boyle et al. (2000) 
quasar luminosity function to calculate the number density of
quasars ($\rho_{act}$) in the three magnitude bins at $z=2$.
For this calculation we adopt the Boyle et al. polynomial fit to 
the luminosity function, converting the absolute magnitude limits and 
number densities to our adopted cosmology. Finally, the estimated
number densities of active black-holes for each of the mass thresholds
is simply the sum of $f\times \rho_{act}$ over the three magnitude bins.

\begin{figure*}
\centerline{\epsfig{file=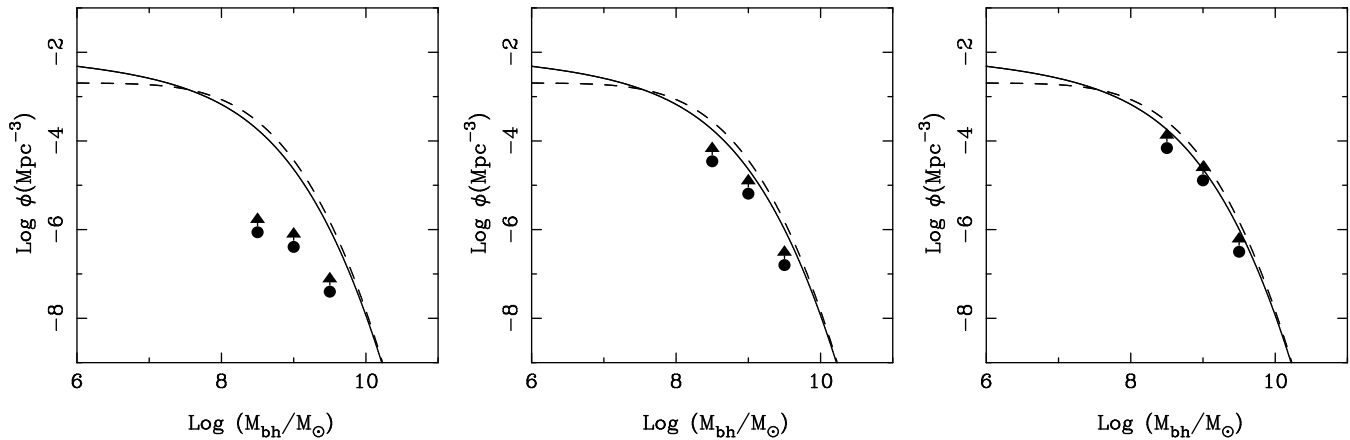,width=18.0cm,angle=0}}
\caption{In each panel the solid and dashed lines show the cumulative
local dormant black-hole mass functions as derived in Section 4 using
the $M_{bh}-L_{bulge}$ and $M_{bh}-\sigma$ relations respectively. In
panel A we show the estimated number densities of active 
supermassive black-holes at $z\simeq 2$ for three mass thresholds
($\geq10^{8}\Msolar, \geq10^{9}\Msolar\, \& \geq10^{9.5}\Msolar$). In 
panel B these number densities have be adjusted to account 
for a possible relationship between mean quasar lifetime 
and black-hole mass (see text for discussion). In Panel C the 
number densities of supermassive black-holes have been 
boosted by a factor of two to conservatively account for geometric
obscuration.}
\label{fig6}
\end{figure*}

The estimates of the number densities of active supermassive black
holes at $z\simeq 2$ using the above method are shown in panel A of 
Fig \ref{fig6}, for the three black-hole mass thresholds. Also shown
are the $z\simeq 0$ cumulative black-hole mass functions based on our
$M_{bh}-L_{bulge}$ and $M_{bh}-\sigma$ estimates from Section 4. 
This plot indicates that the fraction of present-day black holes which
are in place, active, {\it and} optically unobscured at $z \simeq 2$ is
$\simeq 0.005$ for
$M_{bh} \ge 10^{8.5} {\rm M_{\odot}}$,  $\simeq 0.02$ for 
$M_{bh} \ge 10^{9} {\rm M_{\odot}}$, and $\simeq 0.05$ for
$M_{bh} \ge 10^{9.5} {\rm M_{\odot}}$.
 
How should we interpret these ratios? The actual observed number density
ratio at a given mass threshold, R, must be a product of three physical
ratios, namely:
\begin{equation} 
$$R = (n_{z=2}/n_{z = 0}) \times (t_Q/t_{epoch}) \times
(n_{vis}/n_{obs})$$
\end{equation}
\noindent 
where $n_{z=2}$ is the number density of black holes 
(above the appropriate mass threshold) already in place at $z = 2$,
$n_{z=0}$ is the corresponding number density at the present day, $t_Q$
is the net lifetime of a black hole 
(above the appropriate mass threshold) in the optically luminous quasar
phase, $t_{epoch}$ is the total duration of the luminous
quasar epoch, and the ratio $(n_{vis}/n_{obs})$ describes the effect of
geometrical obscuration by dusty tori (note the 
effect of a completely obscured phase is different, and is implicitly
taken care of by the definition of $t_Q$).

Thus, while at first sight it might appear that panel A in Fig \ref{fig6}
indicates that the fraction of present-day black holes already
in place at $z \simeq 2$ is an increasing function of mass, this could
equally well reflect a mass-dependence 
of either net quasar lifetime, or torus opening angle. Some interesting
limiting values can, however, be deduced with reasonable confidence. For
example, adopting $t_{epoch} = 2$ Gyr (corresponding to the redshift
range $1.5 < z < 3.5$, the full width half maximum of the peak in luminous
quasar number density - see, e.g. Richstone et al. (1998); Osmer 2003),
we can infer a lower
limit on the optical quasar lifetime of the most massive black holes $M
\ge 10^{9.5} {\rm M_{\odot}}$ of $t_Q > 1 \times 10^{8}$ yr
if there are no type II quasars, or $t_Q > 2 \times 10^{8}$ yr if
$(n_{vis}/n_{obs}) \simeq 0.5$ as found at the bright end 
of the radio-loud quasar population. Such numbers seem very reasonable,
agreeing well with, for example, the recent findings 
of Yu \& Tremaine (2002). The fact that $t_Q > 1 \times 10^8$ yr differs
spectacularly from, for example, the conclusion reached
by Richstone et al. (1998) that  $t_Q \simeq 1 \times 10^6$ yr, can be
traced to the fact that our inferred quasar black-hole masses are on
average larger (because we have not been forced to assume Eddington
limited accretion) and the fact our 
analysis is confined to the very high-mass end of the black-hole mass
function.
 
In fact, as a result of their analysis, Yu \& Tremaine (2002) find
evidence for a mass dependence of $t_Q$ which can largely explain the
apparent mass dependence of activity ratio shown in panel A of Fig \ref{fig6}.
Specifically, Yu \& Tremaine determined 
that the mean quasar lifetime should be an increasing function of
black-hole mass, rising from $3 \times 10^{7}$ yr at $10^{8}\Msun$
to $5 \times 10^{8}$ yr at  $10^{9.5}\Msun$. For illustrative
purposes, we show in panel B of 
Fig \ref{fig6} the effect of correcting the estimated number densities at $z
\simeq 2$, using the appropriate values of $t_Q$ as 
a function of mass (from Fig. 5 of Yu \& Tremaine) and again adopting
$t_{epoch} = 2$ Gyr. This has the effect of removing 
all significant mass dependence, and raises the apparent black-hole
activation ratio to $\ge 0.2$ for all three mass 
thresholds. Applying a conservative correction of $\simeq 2$  for
geometric obscuration raises this ratio to $\ge 0.4$ (see panel C of
Fig \ref{fig6}). Given 
that our $z \simeq 2$ census is undoubtedly incomplete (being confined
to the very luminous end of the quasar 
population) this lower limit is sufficiently close to unity to imply
that the majority of supermassive black holes are in place
by $z \simeq 2$, and to leave little, if any, room for an additional
substantial population of completely obscured growing black holes at the
high-mass end. This finding is in accord with the conclusions of Yu \&
Tremaine (2002), and Fabian (2003) that the apparently substantial
obscured population of AGN required to explain the X-ray background is
to be found at lower black hole masses than those probed by the present
study.
 
Finally, we note that values of quasar lifetime ($t_Q$) as large as a
significant fraction of a Gyr as discussed above, have often been
rejected as unfeasible for the most massive black holes $M > 10^9
\Msun$. This is because, if accretion is 
assumed to proceed at, or close to, the Eddington limit, extreme
black hole masses $M > 10^{10} \Msun$ will then
be produced in only $1 \times 10^8$ yr, and as discussed in Section 3.2
we find no evidence for any significant population
of such extreme-mass black holes within the SDSS quasar sample. However,
the lack of such extreme objects can be 
reconciled with our inferred lower limit on quasar lifetime $t_Q > 2
\times 10^8$ yr by revisiting the results shown in Fig
\ref{fig3}. Here it can be seen that continued Eddington-limited 
exponential growth towards   $M \simeq 10^{10}
\Msun$ requires mass accretion rates which rapidly approach $100
{\rm M_{\odot} yr^{-1}}$. In contrast, allowing for the known scatter 
in the calculation 
of bolometric luminosities (Elvis et al. 1994), Fig \ref{fig3}
provides good evidence that the most massive black holes are not 
accreting matter at a rate significantly in
excess of  $10 {\rm M_{\odot} yr^{-1}}$. Clearly, a $10^9 \Msun$
black hole can continue to accrete matter at this rate, and produce
bright quasar light for a substantial fraction 
of a Gyr without producing a final black-hole mass in excess of a few
times $10^9 \Msun$. 
 
In conclusion, our results imply that the majority of the most massive
black-holes are in place by $z \simeq 2$ and that the most
massive black holes accrete the bulk of their final mass (i.e. as they
grow from $\simeq 10^{8.5} \Msun$ to $\simeq 10^{9.5} \Msun$)
as optically luminous quasars, at a growth rate limited {\it not} by the
Eddington limit but by some other
physical limit on fuel supply which prevents accretion rates
significantly in excess of $10 {\rm M_{\odot} yr^{-1}}$. Such a physical
limit on black-hole fuel supply might be imposed by accretion disc
physics (e.g. the calculations of 
Burkert \& Silk (2001) indicate that accretion disc viscosity can be
expected to limit the mass consumption rate of a 
supermassive black-hole at the centre of a forming spheroid to $\simeq
2 - 20 {\rm M_{\odot} yr^{-1}}$) or by the physics 
of galaxy formation (e.g. Archibald et al. 2002; Granato et al. 2001,
2003). In addition, one might speculate that the apparent 
lack of a substantial population of highly obscured supermassive black
holes may be a consequence of this sub-Eddington accretion, as compared
to the Eddington-limited accretion more likely to be experienced by
lower mass objects (Fabian 2003).

\section{Conclusions}

Virial black-hole mass estimates have been calculated for a sample of
12698 quasars in the redshift interval $0.1<z<2.1$ drawn from
the SDSS Quasar Catalogue II (Schneider et al. 2003). The distribution
of the quasar black-hole masses as a function of redshift has 
been presented and compared with the masses of dormant 
black-holes observed in the local Universe. In addition, the 
quasar host-galaxy properties implied from an application of the 
locally observed relationship between black-hole and bulge mass have
been compared with the known properties of local early-type galaxies.
By combining the black-hole mass estimates 
with the quasar bolometric luminosities the distribution of quasar
accretion rates have been investigated. Furthermore, in combination with 
the optical quasar luminosity function, the new SDSS black-hole mass 
estimates have been used to estimate the number density of 
{\it active} black-holes at $z\simeq 2$ as a
function of mass. Finally, the activation fraction 
of black holes at $z\simeq 2$ has been estimated by comparing 
the estimated number density of active black holes with
the local dormant black-hole mass function. The main conclusions of
this study can be summarized as follows:

\begin{enumerate}

\item{The virial black-hole mass estimates of the SDSS quasars are
entirely consistent with an upper boundary of $\sim 3\times
10^{9}\Msun$. This limit is consistent with both the most massive
black-holes measured dynamically in the local Universe, and the
expected black-hole mass limit based on the known properties of
early-type galaxies and the locally observed correlation 
between bulge and black-hole mass. Consequently, using 
the MgII-based virial mass estimator, no evidence 
is found for a conflict between quasar black-hole masses in the
redshift range $0.1<z<2.1$ and the contemporary, or ultimate properties
of their host-galaxy population.}

\item{The estimated Eddington ratios of the SDSS quasars show only a small
level of evolution over the redshift range $0.0<z<2.1$, rising from 
$L_{bol}\simeq 0.15 L_{Edd}$ at 
$z\simeq 0.2$ to $L_{bol}\simeq 0.5 L_{Edd}$ at $z\simeq 2.0$.
Although the most luminous broad-line SDSS quasars are accreting 
at rates up to the Eddington limit by $z\simeq 1$, there does not 
appear to be a significant change in accretion rates in 
the redshift range $1<z<2$, with the most luminous broad-line SDSS quasars 
in the current sample still accreting within a factor 
of $\simeq 2$ of the Eddington limit at $z\simeq 2$. Consequently, it 
appears from these results that the Eddington limit is still 
a relevant physical boundary to the accretion rate of luminous 
broad-line quasars, at least at $z\, \ltsim \,2$.}

\item{Using the latest determinations of the galaxy luminosity and
stellar-velocity dispersion functions, estimates of the local
dormant black-hole mass function for early-type galaxies using the
$M_{bh}-\sigma$ and $M_{bh}-L_{bulge}$ relations are shown to 
be consistent at the high-mass end ($M_{bh}>10^{8}\Msun$). The 
agreement between the two methods is found to be
dependent on the adoption of similar levels of association scatter
($\simeq 0.3$ dex) in both correlations. Our best estimate of the
total mass density of dormant black holes within the local early-type
galaxy population is $\rho_{bh}=(2.8\pm 0.4)\times 10^{5} 
\Msolar {\rm Mpc}^{-3}$.}

\item{The activation fraction of supermassive black-holes at $z\simeq
2$ is apparently an increasing function of mass, with an activation
rate of $f\simeq 0.005$ at $M_{bh}\simeq 10^{8.5}\Msun$ rising 
to $f\simeq 0.05$ at $M_{bh}\simeq 10^{9.5}\Msun$. However, it is shown that 
this result is consistent with theoretical work which predicts quasar
lifetimes to be an increasing function of black-hole mass. Correcting
for this expectation we find that the shape of the active 
black-hole mass function at $z\simeq 2$ is consistent with that of 
the local dormant black-hole mass function, with a normalization a 
factor of $\simeq 5$ lower. Making a conservative correction of a
factor of two to account for geometric obscuration, the direct 
implication of this result is that the fraction of black-holes 
with mass $\geq 10^{8.5}\Msun$ which are in place and 
active at $z\simeq 2$ is $\geq 0.4$.}

\item{A fairly robust limit on the lifetime of quasars with black-hole
masses $\geq 10^{9.5}\Msun$ is found to be $t_Q>2\times 10^{8}$
years. Black holes of this mass appear to be prevented from growing to
masses $>10^{10}\Msun$ by some physical mechanism, other than the
Eddington limit, which prevents accretion at rates $\gtsim\, 10\,\Msun$
per year.}

\end{enumerate}

\section{acknowledgments}
RJM acknowledges the award of a PPARC PDRA. 
JSD acknowledges the enhanced research time provided by the
award of a PPARC Senior Fellowship. The authors acknowledge 
Matt Jarvis for useful discussions. Funding for the 
Sloan Digital Sky Survey (SDSS) has been provided by the 
Alfred P. Sloan Foundation, the Participating Institutions, 
the National Aeronautics and Space Administration, the National 
Science Foundation, the U.S. Department of Energy, the 
Japanese Monbukagakusho, and the Max Planck Society. The 
SDSS Web site is http://www.sdss.org/. The SDSS is managed by the 
Astrophysical Research Consortium (ARC) for the Participating 
Institutions. The Participating Institutions are The University of 
Chicago, Fermilab, the Institute for Advanced Study, the Japan 
Participation Group, The Johns Hopkins University, Los Alamos 
National Laboratory, the Max-Planck-Institute for Astronomy (MPIA), 
the Max-Planck-Institute for Astrophysics (MPA), New Mexico State 
University, University of Pittsburgh, Princeton University, the 
United States Naval Observatory, and the University of
Washington.

\begin{appendix}
\section{The virial black-hole mass estimate}
\label{virial}

The assumption that certain broad-emission lines
present in the spectra of Type 1 quasars trace the virialized velocity
field of the broad-line region (BLR) leads to the so-called virial
black-hole mass estimate:
\begin{equation}
M_{bh}=G^{-1}R_{BLR}V_{BLR}^{2}
\label{mass}
\end{equation}
\noindent
where $R_{BLR}$ is the BLR radius and $V_{BLR}$ is the
Keplerian velocity of the line-emitting gas. Recent evidence in the
literature which supports the basic assumption of virialized motions
in the BLR has been previously discussed in introduction. In this
Appendix we describe the derivation of the final versions of the virial
mass estimators, based on both the $H\beta$ and MgII emission lines,
used to estimate the SDSS quasar black-hole masses. Much of this
material has been discussed previously by McLure \& Jarvis (2002),
and the reader is referred to this paper for more details. The material
is partly repeated here for completeness, and to highlight a change in
the adopted calibration of the $R_{BLR}-\lambda L_{\lambda}$ relation 
at 3000\,\AA\,which is integral to the MgII based virial 
black-hole mass estimator.

\subsection{Estimating the broad-line region radius}
\label{radius}

In the absence of reverberation mapping it is necessary to estimate
$R_{BLR}$ via the correlation between $R_{BLR}$ and AGN continuum
luminosity (Wandel, Peterson \& Malkan 1999; Kaspi et al. 2000). 
In McLure \& Jarvis (2002) the best-fitting $R_{BLR}-\lambda
L_{\lambda}$ relations were derived at both 3000\,\AA\, and
5100\,\AA\, for the
34 AGN with reverberation mapping estimates of $R_{BLR}$ from Wandel,
Peterson \& Malkan (1999) and Kaspi et al. (2000). Using the BCES 
method of Akritas \& Bershady (1996) the bisector fits 
at both wavelengths were as follows:
\begin{equation}
R_{BLR}=(25.2\pm3.0) \left[\lambda L_{3000}/10^{37}W
\right]^{(0.47\pm0.05)}
\end{equation}
\begin{equation}
R_{BLR}=(26.4\pm4.4) \left[\lambda L_{5100}/10^{37}W
\right]^{(0.61\pm0.10)}
\end{equation}
\noindent
where $R_{BLR}$ is in units of light-days. The sample of 34 AGN with
reverberation mapping estimates of $R_{BLR}$ is composed of 17 PG
quasars and 17 lower-luminosity Seyfert galaxies spanning the 
luminosity range $10^{34}$W$<\lambda L_{\lambda}<10^{39}$W. While 
this luminosity range is fairly
representative of the SDSS quasars analysed using
the $H\beta$ emission line ($0.1<z<0.5$), it is not representative of
the SDSS quasars at $z\ge0.7$, where we have utilized the MgII
emission line to estimate their black-hole masses. Simply as a result 
of their higher redshifts, these quasars reside exclusively in the 
luminosity range 
$10^{37}$W$<\lambda L_{\lambda}<10^{40}$W. Consequently, in order to
provide as accurate an estimate of the $R_{BLR}$ for each quasar as
possible, it was decided to re-fit the $R_{BLR}-\lambda
L_{\lambda}$ relation at 3000\,\AA\, using only those objects from the
reverberation mapped AGN sample with $\lambda L_{\lambda}
>10^{37}$W. Following McLure \& Jarvis (2002) we again adopt the BCES
bi-sector fit as our best estimate of the $R_{BLR}-\lambda
L_{\lambda}$ relation at 3000\,\AA\,, with the following result:
\begin{equation}
R_{BLR}=(18.5\pm6.6) \left[\lambda L_{3000}/10^{37}W
\right]^{(0.62\pm0.14)}
\end{equation}
\noindent
which is shown along with the previous $R_{BLR}-\lambda
L_{3000}$ fit from McLure \& Jarvis (2002) in Fig A1. It can be seen from
Fig A1 that although the two fits are consistent, the new
steeper $R_{BLR}-\lambda L_{3000}$ relation does provide a better fit
to the reverberation mapped PG quasars of Kaspi et al. with
$L_{\lambda}>10^{37}$W, which are typical of the SDSS quasars 
at $z\geq 0.7$. Consequently, it is the new steeper form of the
$R_{BLR}-\lambda L_{3000}$ relation which is adopted throughout the
analysis performed in this paper. 
\footnote{The mean difference in black-hole mass introduced by 
using Eqn A4 in preference to Eqn A2 is only 0.086 dex and does not
significantly change any of the results or conclusions in this paper.} 
The fit to the $R_{BLR}-\lambda L_{5100}$ relation determined 
by McLure \& Jarvis (2002) is adopted unchanged because of the 
lower luminosity range occupied by the SDSS quasars at $z\leq 0.7$.
\begin{figure}
\centerline{\epsfig{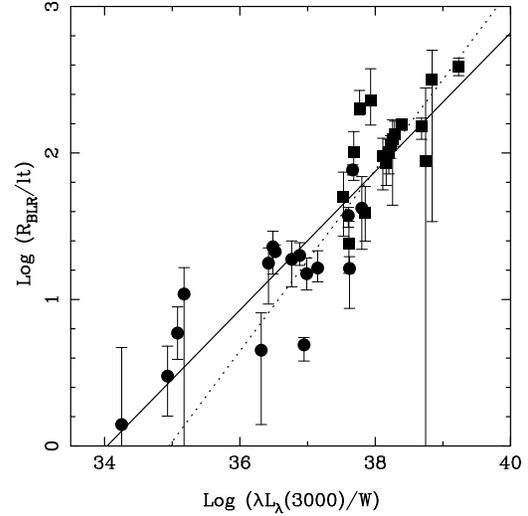}}
\caption{The $R_{BLR}-\lambda L_{3000}$ relation for the 34 AGN with
reverberation mapping estimates of $R_{BLR}$ from Kaspi et
al. (2000). The solid line is the best-fitting BCES bi-sector fit
determined for the full sample by McLure \& Jarvis (2002); Eqn A2. 
The dotted line shows the new steeper relation adopted in this paper, 
which is a fit to those objects with $\lambda L_{\lambda}>10^{37}$W.}
\end{figure}

\subsection{Broad-line region geometry}
In Eqn \ref{mass} the velocity of the BLR gas is 
taken as $V_{BLR}=$~$f\times$~FWHM, where FWHM is the full-width half
maximum of your chosen broad emission line (either $H\beta$ or MgII in
this case) and $f$ is a geometric factor which relates the FWHM to 
the intrinsic Keplerian velocity of the broad-line gas.

Due to the fact that there is currently no consensus on the 
geometry of the BLR in radio-quiet quasars it is conventional to set
$f=\sqrt{3}/2$ , which is appropriate if the orbits of the BLR gas are
randomly orientated. Perhaps the most likely alternative to randomly
orientated BLR orbits is the possibility that the geometry of the BLR
is flattened or disk-like in nature. In radio-loud quasars, where a
rough orientation indicator is available from the radio core-to-lobe
ratio, there is good evidence that broad-emission line FWHM show a
orientation dependence consistent with a disk-like geometry
(e.g. Browne \& Wills 1986; Brotherton 1996). Given the similarity
between the optical emission-line spectra of radio-loud and
radio-quiet quasars, it is perhaps not unreasonable to consider the
possibility that the BLR of radio-quiet quasars (which dominate the
SDSS sample) are also flattened. As discussed by McLure \& Dunlop
(2002), in this case: 
\begin{equation}
V_{BLR}=FWHM/(2\times \sin i) 
\end{equation}
\noindent
where $i$ is the angle between the line-of-sight and the disk axis 
(i.e. $i=0\degmark$ is pole-on). It has been established for
radio-loud quasars that the angle to the line-of-sight is consistent 
with being randomly distributed between 
$0\degmark<i<45\degmark$ (Barthel 1989; Willott et
al. 2000). Assuming the same situation to be true for radio-quiet
quasars, it follows that the mean line-of-sight angle will be
$<i>=30\degmark$ (averaging over solid angle) implying that
$<f>=1$. Consequently, it can be seen that $f=1$ represents a good
choice of geometric factor in the case of both random or flattened BLR
orbits and is adopted throughout this paper. We note that if the BLR
of radio-quiet quasars is flattened, objects which are aligned closer
than 15$\degmark$ to the line-of-sight will have their black-hole
masses underestimated by a factor of $>4$. However, these objects 
should only represent a small fraction of the SDSS 
sample ($\simeq 10\%$ if $i$ is randomly distributed in the range
$0\degmark<i<45\degmark$) and will likely appear as
narrow-line objects (FWHM$<2000$ km s$^{-1}$), which are not the principal
focus of this paper. A discussion of the evidence for the flattened
BLR hypothesis provided by the SDSS sample is presented in Section 3.3.

\subsection{The virial black-hole mass estimators}
Substituting the calibrations of the  $R_{BLR}-\lambda
L_{\lambda}$ relations into Eqn A1, and making the
identification that $V_{BLR}$ = MgII or H$\beta$ FWHM, we arrive at the
virial black-hole mass estimators. As in McLure \& Jarvis (2002), these 
virial estimators are then compared to the black-hole masses of 
the reverberation mapped AGN from the Kaspi et al. sample. 
By requiring that on average the new virial estimators match 
the mass estimates derived by Kaspi et al. (2000), using the 
reverberation mapping values of $R_{BLR}$ and rms FWHM measurements,
small off-sets of $-0.05$ dex and $-0.04$ dex are introduced to 
the MgII and H$\beta$ estimators respectively. 
Accounting for these off-sets
we arrive at the final versions of the virial black-hole mass estimators
adopted throughout the analysis in this paper:
\begin{equation}
\frac{ M_{bh}} {\Msolar}  =3.2\left(\frac{\lambda
L_{3000}}{10^{37}{\rm W}}\right)^{0.62}\left(\frac{FWHM(MgII)}
{{\rm kms}^{-1}}\right)^{2}
\label{final}
\end{equation}
\begin{equation}
\frac{M_{bh}}{\Msolar}=4.7\left(\frac{\lambda
L_{5100}}{10^{37}{\rm W}}\right)^{0.61}\left(\frac{FWHM(H\beta)}
{{\rm kms}^{-1}}\right)^{2}
\label{optical}
\end{equation}
\noindent
where the H$\beta$ based black-hole mass estimator is unchanged from
that derived by McLure \& Jarvis (2002).

\subsection{Comparing the ${\bf H\beta}$ and MgII based estimators}

\begin{figure}
\centerline{\epsfig{file=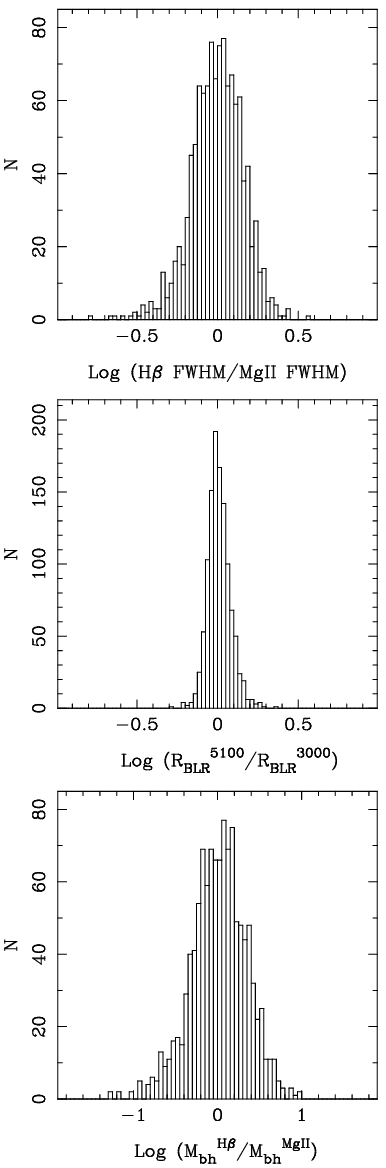,width=6.cm,height=19cm,angle=0}}
\caption{The top panel shows a histogram of the log of the ratio 
of H$\beta$/MgII FWHM for 1136 objects in the redshift 
interval $0.5<z<0.8$. The mean log ratio is $-0.004\pm0.005$, with a
dispersion of 0.16 dex. The middle panel shows a histogram of 
the log of the ratio of $R_{BLR}$ as estimated
via the  $R_{BLR}-\lambda L_{\lambda}$ relation at 5100\,\AA\, and
3000\,\AA\, respectively. The mean log $R_{BLR}$ ratio is 
$0.0010\pm0.0021$, with a dispersion of 0.07 dex. The bottom panel 
shows a histogram of the log of the ratio of virial black-hole mass
estimate based on the $H\beta$ and MgII emission lines and the 
$R_{BLR}-\lambda L_{\lambda}$ relations at 5100\AA\, and
3000\AA\, respectively. The mean log ratio is $0.013\pm0.010$ with a
dispersion of 0.33 dex.}
\end{figure}

McLure \& Jarvis (2002) compared the $H\beta$ and MgII based virial
estimators using a sample of 150 quasars drawn from the LBQS (Forster
et al. 2001) and MQS samples (Baker et al. 1999) which had 
spectra covering both emission lines. The result of
this comparison showed that $M_{bh}(H\beta) \propto 
M_{bh}(MgII)^{1.00\pm 0.08}$ with a $1\sigma$ scatter in $\log M_{bh}$
of 0.41 dex. Here we can re-investigate the agreement between the two
virial mass estimators more thoroughly using the 1136 SDSS quasars
from our sample in the redshift range $0.5<z<0.8$ where it was
possible to estimate the black-hole mass using both $H\beta$
and MgII. The results of this comparison are
illustrated in Fig A2. The top panel shows the distribution of the
ratio of the FWHMs of the $H\beta$ and MgII emission lines. The 
middle panel shows the distribution of the ratio of broad-line region
radius, as estimated via the $R_{BLR}-\lambda L_{\lambda}$ correlations
at 5100\,\AA\, and 3000\,\AA\, respectively. The distributions of the
FWHM and $R_{BLR}$ ratios shown in the top and middle panels are both
centred on unity, with $1\sigma$ scatters of 0.16 dex and 0.07 dex
respectively. The bottom panel shows the distribution of the ratio of
black-hole mass, as estimated via Eqn A7 and Eqn A6 respectively. The
mean log ratio of the black-hole mass estimates is $0.013$ with a
$1\sigma$ scatter of 0.33 dex. In conclusion, this comparison
demonstrates that the MgII and $H\beta$ based virial mass estimators
return the same mass estimates within a factor of $\simeq 2$ on
average, and that statistically are entirely interchangeable. This
conclusion is re-enforced by the smooth transition between the two
estimators at $z\simeq 0.7$ shown in Fig 1.

\section{Model fitting}

\begin{figure*}
\centerline{\epsfig{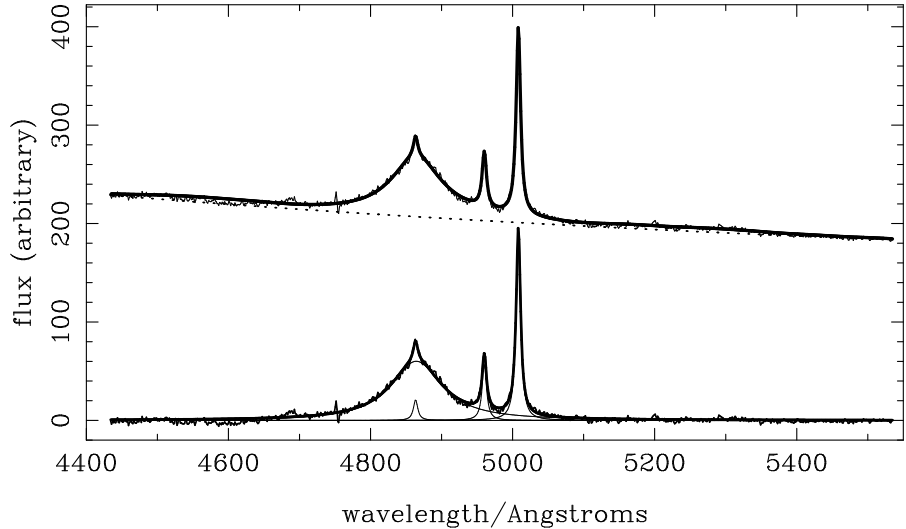}}
\caption{An example of the emission-line fitting process in the $H\beta$
region for an SDSS quasar at z=0.17. The upper plot shows the original
spectrum shifted to the rest-frame (thin line) with the best-fitting
model over-plotted as the thick line. Also shown in the upper plot is
the fitted power-law quasar continuum (dotted line). The lower plot
shows the original spectrum after subtraction of the best-fitting
quasar continuum and iron-emission template. The thick solid line
shows the best-fitting model, while the thin solid line shows the
individual emission-line components which comprise the best-fitting model.}
\end{figure*}

\begin{figure*}
\centerline{\epsfig{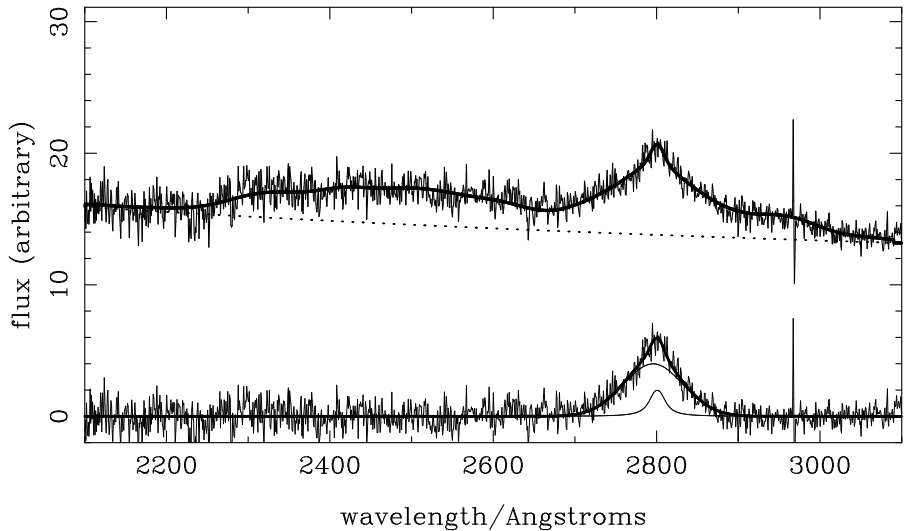}}
\caption{An example of the emission-line fitting process in the MgII 
region for an SDSS quasar at z=0.88. The upper plot shows the original
spectrum shifted to the rest-frame (thin line) with the best-fitting
model over-plotted as the thick line. Also shown in the upper plot is
the fitted power-law quasar continuum (dotted line). The lower plot
shows the original spectrum after subtraction of the best-fitting
quasar continuum and iron-emission template. The thick solid line
shows the best-fitting model, while the thin solid line shows the
individual emission-line components which comprise the best-fitting model.}
\end{figure*}

Due to the size of the sample of SDSS quasars
analysed here, the fitting of the $H\beta$ and MgII emission lines and
the determination of the continuum luminosities at 5100\,\AA\, and
3000\,\AA\, was undertaken in a fully automated fashion. The various
stages in the emission-line modelling of both $H\beta$ and MgII are
discussed below.

\subsection{Continuum and iron-template fitting}

In order to reliably measure the line-widths of the $H\beta$ and MgII
emission lines it is necessary to subtract the surrounding iron
emission. This subtraction process can often be important for 
fitting the $H\beta$ emission line, but is crucial for fitting the
MgII emission line where the wings are usually strongly blended 
with the surrounding FeII emission. The approach adopted here for 
dealing with the iron emission is a variant 
on the technique developed by Boroson \& Green (1992), which is based 
on the fitting of an iron-emission template derived from the 
spectrum of narrow-line Seyfert
galaxy (NLS1) Izw1. As a member of the NLS1 class, Izw1 is ideal for
this purpose because it displays extremely strong iron emission around
both $H\beta$ and MgII with a FWHM of $\simeq 900$
km s$^{-1}$. Consequently, the strength and FWHM of an iron template 
based on Izw1 can be adjusted to match the iron emission of broad-line
quasars where the FWHM of the iron emission is $\geq 2000$ km s$^{-1}$.
The iron emission templates used in the modelling are taken from two
different sources, but are both derived from spectra of Izw1. For the
modelling of the $H\beta$ emission line we have used the optical FeII
emission template derived by Marziani et al. (2003), which is made
freely available by the authors. For the modelling of the MgII
emission line we have reconstructed the FeII+FeIII emission
template derived by Vestergaard \& Wilkes (2001) from the data
published within that paper.

The first stage in the modelling process of both the $H\beta$ and MgII
emission lines is the simultaneous fitting of the underlying quasar
continuum and the iron-emission template, where the quasar continuum
is assumed to be a simple power-law ($f_{\nu} \propto \nu^{-\alpha}$). 
For both emission lines this fitting was performed in two wavelength 
windows, which conservatively
excluded the $H\beta$ and MgII emission regions to avoid biasing the
iron-emission fit with the wings of the line emission. For the
$H\beta$ emission line the continuum+FeII fitting windows were
4435\,\AA\,$\rightarrow$ 4700\,\AA\, and
5100\,\AA\,$\rightarrow$5535\,\AA\,. The equivalent regions for the
MgII emission line where 2100\,\AA\,$\rightarrow$ 2675\,\AA\, and 
2925\,\AA\,$\rightarrow$3090\,\AA\,. During the fitting process the
normalization and slope of the continuum were left as free
parameters, as were the luminosity and FWHM of the iron template. Upon
completion of this process the best-fitting continuum+iron combination 
was subtracted from the spectra, leaving the isolated MgII or
$H\beta$+[O{\sc iii}] emission lines. The fitting of the MgII and
$H\beta$ emission lines then proceeded in slightly different fashions,
and are therefore described separately below.

\subsection{Fitting the $H\beta$+[O{\sc iii}] emission-line complex}

The $H\beta$ and [O{\sc iii}] emission lines were modelled
simultaneously as a combination of four components: broad and narrow
$H\beta$ plus narrow components for [O{\sc iii}] 4959\,\AA\, and 
[O{\sc iii}] 5007\,\AA\,. In order to limit the degrees of
freedom involved in the fitting several restrictions were invoked. 
Firstly, the central wavelengths of the
narrow $H\beta$ component and the [O{\sc iii}] 4959\,\AA\, component were
constrained to lie at their laboratory wavelength offsets with respect
to [O{\sc iii}] 5007\,\AA. This meant that there were only two free
central wavelengths involved in the fit, that of the broad $H\beta$
component and [O{\sc iii}] 5007\,\AA\,. Secondly, the FWHM of the narrow
$H\beta$ component and the two [O{\sc iii}] components were constrained
to have the same value. The final restriction imposed was that the
FWHM of the narrow components was constrained to be less 
than 2000~km s$^{-1}$. 

Four different representations of the line profiles were tried for
each object. Firstly, two fits were performed where all the line profiles
(broad+narrow $H\beta$ and [O{\sc iii}] 4959 \& 5007) were first
modelled as gaussians and then as lorentzians.  Secondly, two further
fits were performed where the broad $H\beta$ profile was first
modelled as a gaussian, with all the narrow profiles modelled as
lorentzians, and then secondly with broad $H\beta$ modelled as a
lorentzian with all the narrow profiles modelled as gaussians. 
The FWHM value used in the virial mass estimates is that of the
best-fitting model profile, in terms of $\chi^{2}$, within the 
line-fitting region (4700\,\AA\,$\rightarrow$ 5100\,\AA\,).

\subsection{Fitting the MgII emission line}

The first stage of the modelling of the MgII emission line involved
the fitting of a two-component, broad+narrow profile. This proceeded 
in similar fashion to the modelling of the $H\beta$ emission line,
with both the narrow and broad components first represented as either
gaussians or lorentzians. In addition, two further model fits were
tried for each object, the first comprising a gaussian broad component
and a lorentzian narrow component, and the second a lorentzian broad
component and a gaussian narrow component. During this stage of the
modelling the central wavelength, FWHM and flux of both the narrow and
broad components were treated as free parameters. The only restriction
invoked was that the narrow component had a FWHM $\leq 2000$
km s$^{-1}$. Both the narrow and broad components were modelled as
doublets separated by their laboratory wavelength difference, and all
of the emission-line models were fitted within the wavelength range
2700\,\AA\,$\rightarrow$ 2900\,\AA\,.

Unlike the $H\beta$ emission line, which often displays a clear
point of inflection between the narrow and broad components, with MgII
it is often unclear whether a significant, separate narrow 
component is present. To account for this, the MgII emission line 
of each quasar was also modelled as a single broad lorentzian or 
gaussian doublet. For the purposes of the virial mass estimates the FWHM of 
the broad component of the best-fitting two-component model profile was 
adopted as the best FWHM estimate for each quasar, provided that two
criteria were satisfied. Firstly, the $\Delta \chi^{2}$ between the
best-fitting two-component and best-fitting single-component models
was required to be significant at the $3\sigma$-level, i.e. the 
extra degrees of freedom in the two-component model had to be justified 
statistically. Secondly, the equivalent width (EW) of 
the broad component was required to be greater than two-thirds 
of the total MgII EW, ensuring that the FWHM of the 
broad component was only adopted if it constituted the dominant 
fraction of the line flux. If either of these two criteria were not
satisfied the FWHM of the best-fitting single-component model profile 
was adopted. The typical uncertainty in the fitted FWHM values for
both the H$\beta$ and MgII emission lines is $\pm15\%$.

\section{Luminosity calibration}

\begin{figure}
\centerline{\epsfig{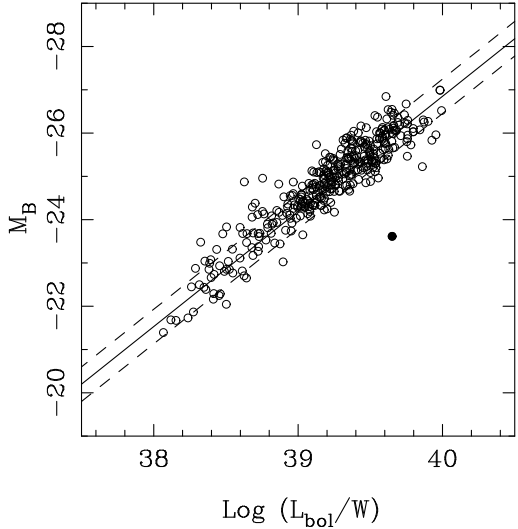}}
\caption{The calibration (solid line) between absolute $B-$band magnitude and
estimated bolometric luminosity, based on N=372 quasars common to 
both the SDSS Quasar Catalog II and the 2dF 10K quasar catalogs. The two 
dashed lines delineate the $1\sigma=0.14$ scatter around the
best-fit. The outlying object shown as a filled circle is
suspected of having an incorrect magnitude in the 2dF 10K quasar 
catalogue (see text for discussion).}
\end{figure}
\noindent

For the purposes of the analysis carried out within the paper it was 
necessary to provide a consistent estimate of the quasar bolometric
luminosities, based on the continuum close to either the $H\beta$
or MgII emission lines. Furthermore, in order to perform the estimate
of the quasar activation rates in Section 5, it was necessary to
provide a calibration between the quasar bolometric
luminosities estimated via the spectra modelling, and the absolute
$B-$band magnitudes used in the Boyle et al. (2000) determination of
the 2dF+LBQS quasar luminosity function. Both of these calibration
issues are discussed below.

\subsection{Estimating the bolometric luminosity}

The bolometric luminosities of each of the SDSS quasars are estimated
from their monochromatic $\lambda L_{\lambda}$ luminosity at either
5100\,\AA\, or 3000\,\AA\,, depending on their redshift. The appropriate 
bolometric correction factors have been estimated by extrapolating
from the bolometric correction at 2500\,\AA\, determined by Elvis et
al. (1994), employing their median quasar SED. Taking into account the
recent re-evaluation of the average 2500\,\AA $\rightarrow$ 2 keV
X-ray slope (Elvis, Risaliti \& Zamorani 2002) we estimate the 
bolometric correction factors to be 5.9 and 11.3 at 
3000\,\AA\, and 5100\,\AA\,respectively. 

Due to the fact that the SDSS sample of quasars analysed here includes
1136 objects in the redshift range $0.5<z<0.8$, whose spectra 
cover both 3000\,\AA\, and 5100\,\AA\,, we have an opportunity to
verify that our choice of bolometric correction factors lead to 
consistent bolometric luminosity estimates at both wavelengths. 
Using correction factors of 5.9 and 11.3 we find
that the mean log ratio of bolometric 
luminosity estimates is : 
$<\log (L_{bol}(5100)/L_{bol}(3000))>=0.062\pm0.003$. This
result is inconsistent with the expected ratio of unity, and is 
equivalent to the 5100\,\AA\, based bolometric luminosity estimate 
being $15\%$ greater on average. 
Consequently, we chose to
revise downward the 5100\,\AA\, bolometric correction factor to 9.8,
which ensures a mean ratio of unity for the 1136 SDSS quasars whose 
spectra cover both 5100\,\AA\, and 3000\,\AA\,, with a $1\sigma$ 
scatter of less than $30\%$. 

\subsection{The ${\bf M_{B}}$ - bolometric luminosity calibration}

The calibration between absolute $B-$band magnitude and the
bolometric luminosities derived from the fits to the flux calibrated
SDSS spectra was done using a sample of 372 objects common to the 2dF 
10K quasar catalogue (Croom et al. 2001) and the new SDSS Quasar Catalogue II
(Schneider et al. 2003). The absolute $B-$band magnitudes
for each quasar have been calculated from the apparent  $B_{j}$
magnitudes listed in the 2dF 10K quasar catalog, using the K-corrections and
quasar colours from Cristiani \& Vio (1990) for consistency 
with Boyle et al. (2000). The best-fitting relation to the correlation
between bolometric luminosity and absolute $B-$band magnitude is:
\begin{equation}
M_{B}=-2.66(\pm 0.05)\log \left[L_{bol}/W\right] + 79.36(\pm 1.98)
\end{equation}
\noindent
which is shown in Fig C1. The $1\sigma$ scatter at a fixed absolute
magnitude is 0.14 dex, as highlighted by the
dashed lines. Three of the 372 objects were found to be large outliers and 
were excluded from the fitting process. The most extreme
outlier (shown as the filled circle in Fig C1) is the $z=1.296$ quasar
SDSS J133855.04-010229.9 (2QZ J133855.0-010230). In this case, the 
source of the discrepancy appears to be an 
erroneous $B_{j}$ magnitude listing in
the 2dF 10K catalogue. The $B_{j}$ magnitude for this quasar is
listed as $B_{j}=20.82$. However, the SDSS photometry for this object
finds a $g-$band magnitude of 18.42 with a $g-r$ colour of 0.18. Even
accounting for the switch between AB/Vega magnitudes and galactic
extinction, it is clear that the SDSS and 2dF photometry for this
object are inconsistent, whereas the SDSS PSF magnitude and flux
calibrated spectra are consistent.

\end{appendix}
\end{document}